\documentclass{ws-procs9x6}

\begin{document}

\title{Equivalence Postulate and the Quantum Potential of Two Free Particles}

\author{MARCO MATONE}

\address{Department of Physics ``G. Galilei'' -- Istituto Nazionale di
Fisica Nucleare \\
University of Padova, Italy}


\maketitle

\abstracts{Commutativity of the diagram of the maps connecting
three one--particle state, implied by the Equivalence Postulate
(EP), gives a cocycle condition which unequivocally leads to the
quantum Hamilton--Jacobi equation. Energy quantization is a direct
consequences of the local homeomorphicity of the trivializing map.
We review the EP and show that the quantum potential for two free
particles, which depends on constants which may have a geometrical
interpretation, plays the role of interaction term that admits
solutions which do not vanish in the classical limit.}

\section{The Equivalence Postulate}

Let us consider two one--dimensional one--particle state. The
Equivalence Postulate\cite{1} (EP) is the condition that the
coordinate transformation
\begin{equation} \mathcal{S}_0^b(q^b)=\mathcal{S}_0^a(q^a),
\label{abinitio}\end{equation} be well--defined for {\it any} pair
of states. $\mathcal{S}_0$ is also characterized by the condition
that in a suitable limit reduces to the Hamiltonian characteristic
function (also called reduced action) $\mathcal{S}_0^{cl}$.
Eq.(\ref{abinitio}) implies that ($\mathcal{W}(q)\equiv V(q)-E$)
\begin{equation}\mathcal{W}^a(q^a)\longrightarrow
\mathcal{W}^b(q^b)=
\left(\partial_{q^b}q^a\right)^2\mathcal{W}^a(q^a)+(q^a;q^b),
\label{ladoppiaww}\end{equation} where, due to the commutative
diagram of maps
\begin{equation}
\matrix{ & & b& & \cr & \nearrow & & \searrow & \cr a& &
\longrightarrow & & c}
\label{ildiagrammacommutativo}\end{equation} the unknown term
$(q^a;q^b)$ must satisfy the cocycle condition\cite{1}
\begin{equation}
(q^a;q^c)=\left(\partial_{q^c}q^b\right)^2(q^a;q^b)+(q^b;q^c).
\label{inhomtrans}\end{equation} It is well--known that this is
satisfied by the Schwarzian derivative. However, it turns out that
it is essentially the unique solution, that is we have\cite{1}

\vspace{.333cm}

\noindent {\bf Theorem 1.} {\it Eq.}(\ref{inhomtrans}) {\it
defines the Schwarzian derivative up to a multiplicative constant
and a coboundary term.}

\vspace{.333cm}

\noindent Since the differential equation for $\mathcal{S}_0$
should depend only on $\partial_q^k\mathcal{S}_0$, $k\geq1$, it
follows that the coboundary term must be zero,\cite{1} so that
\begin{equation}
(q^a;q^b)=-{\beta^2\over4m}\{q^a,q^b\},
\label{laderivataschwarziana}\end{equation} where
$\{f(q),q\}=f'''/f'-3(f''/f')^2/2$ is the Schwarzian derivative
and $\beta$ is a nonvanishing constant that we identify with
$\hbar$. As a consequence, $\mathcal{S}_0$ satisfies the Quantum
Stationary Hamilton--Jacobi Equation (QSHJE)\cite{1}
\begin{equation} {1\over
2m}\left({\partial\mathcal{S}_0(q)\over\partial q}\right)^2+V(q)-E
+{\hbar^2\over4m}\{\mathcal{S}_0,q\}=0. \label{1Q}\end{equation}
$\psi={\mathcal{S}_0'}^{-1/2}\left(A
e^{-{i\over\hbar}\mathcal{S}_0}+Be^{{i\over\hbar}\mathcal{S}_0}\right)$
solves the Schr\"odinger Equation (SE)
\begin{equation} \left(-{\hbar^2\over2m}{\partial^2\over\partial
q^2}+V\right)\psi=E\psi.\label{yz1xxxx4}\end{equation} The ratio
$w=\psi^D/\psi$ of two real linearly independent solutions of
(\ref{yz1xxxx4}) is, in deep analogy with uniformization theory,
the {\it trivializing map} transforming any $\mathcal{W}$ to
$\mathcal{W}^0\equiv0$. This formulation, proposed in
collaboration with Faraggi, extends to higher dimension and to the
relativistic case as well.\cite{1}

Let $q_{-/+}$ be the lowest/highest $q$ for which $\mathcal{W}(q)$
changes sign, we have\cite{1}

\vspace{.333cm}

\noindent {\bf Theorem 2.} {\it If} \begin{equation}
V(q)-E\geq\left\{\begin{array}{ll}P_-^2
>0,&q<q_-,\\ P_+^2 >0,&q> q_+,\end{array}\right.
\label{perintroasintoticopiumeno}\end{equation} {\it then $w$ is a
local self--homeomorphism of $\hat{\bf R}$ iff
Eq.}(\ref{yz1xxxx4}) {\it has an $L^2(\bf R)$ solution.}

\vspace{.333cm}

\noindent Since the QSHJE is defined iff $w$ is a local
self--homeomorphism of $\hat{\bf R}$, it follows that energy
quantization is implied by the QSHJE itself.

\noindent Let us now review the derivation of
Eqs.(\ref{abinitio})--(\ref{1Q}). We first look for a coordinate
transformation identifying two classical one--dimensional
one--particle state. We could tentatively impose the apparently
harmless condition for the classical reduced action to transform
as a scalar
\begin{equation}\mathcal{S}^{cl\, b}_0(q^b)= \mathcal{S}^{cl\,
a}_0(q^a),\label{ilprimostep}\end{equation} so that
$q^a\rightarrow q^b= {\mathcal{S}^{cl\,
b}_0}^{-1}\circ\mathcal{S}^{cl\, a}_0(q^a)$. However, there is an
inconsistency which arises for {\it all} possible pairs of states.
In particular, since for a free particle of vanishing energy the
reduced action is a constant, it follows that even if the
transformation is well--defined, the same becomes inconsistent
once the two actions are described in a frame which is at rest
with respect to one of them. Hence, in Classical Mechanics (CM),
the equivalence under coordinate transformations requires choosing
a frame in which no particle is at rest. In order to make
coordinate equivalence a frame independent concept, we postulate
that Eq.(\ref{abinitio}) is always defined. That is given two
one--particle state with reduced actions $\mathcal{S}_0$ and
$\mathcal{S}_0^v$, it always exists the ``$v$--map'' $q\rightarrow
q^v$ defined by\cite{1} $\mathcal{S}^v_0(q^v)=\mathcal{S}_0(q)$.
Let ${\bf H}$ be the space of all possible $\mathcal{W}$. This
condition is essentially the same of imposing the EP\cite{1}

\vspace{.333cm}

\noindent {\it For each pair $\mathcal{W}^a,\mathcal{W}^b\in{\bf
H}$, there exists a $v$--transformation such that}
\begin{equation}
\mathcal{W}^a(q)\longrightarrow{\mathcal{W}^a}^v(q^v)=\mathcal{W}^b(q^v).
\label{equivalence}\end{equation}

\vspace{.333cm}

The fact that one point cannot be diffeomorphic to a line is
essentially the reason why the EP excludes the existence of states
corresponding to a point in phase space; that is the EP implies a
sort of nonlocalization in phase space which is reminiscent of the
Heisenberg uncertainty relation. Another way to see that the EP
cannot be implemented in CM is to note that the two Classical
Stationary HJ Equations (CSHJE)
\begin{equation}
{1\over2m}\left({\partial\mathcal{S}^{cl\, a}_0(q^a)\over\partial
q^a} \right)^2+\mathcal{W}^a(q^a)=0,\quad
{1\over2m}\left({\partial\mathcal{S}^{cl\, b}_0(q^b)\over\partial
q^b} \right)^2+\mathcal{W}^b(q^b)=0,
\label{CSHJEDabinitio}\end{equation} and Eq.(\ref{ilprimostep})
give $\mathcal{W}^b(q^b)=(\partial_{q^b}q^a)^2\mathcal{W}^a(q^a)$.
It follows that the state corresponding to $\mathcal{W}^0\equiv 0$
is a fixed a point in $\bf H$, that is $\mathcal{W}^0\to
(\partial_{q^v}q)^2\mathcal{W}^0\equiv0$. The only way to
eliminate this fixed point is to admit an inhomogeneous term in
the transformation properties of $\mathcal{W}$, as in
Eq.(\ref{ladoppiaww}). On the other hand, we saw that the
transformation properties of $\mathcal{W}$ are fixed by the CSHJE,
so to make (\ref{abinitio}) consistent with (\ref{ladoppiaww}), we
must modify the CSHJE by adding a term $Q(q)$ that for the time
being is completely arbitrary, that is
\begin{equation}
{1\over2m}\left({\partial\mathcal{S}_0(q)\over\partial
q}\right)^2+\mathcal{W}(q)+Q(q)=0.
\label{aa10bbbxxxbprima}\end{equation}
Eqs.(\ref{abinitio})(\ref{ladoppiaww}) and
(\ref{aa10bbbxxxbprima}) give
$\mathcal{W}^b(q^b)+Q^b(q^b)=(\partial_{q^b}q^a)^2[\mathcal{W}^a(q^a)+Q^a(q^a)]$,
and \begin{equation} Q^a(q^a)\longrightarrow
Q^b(q^b)=\left(\partial_{q^b}q^a\right)^2Q^a(q^a)-(q^a;q^b).
\label{laQQQ9}\end{equation}  The main steps in deriving theorem 1
are the two lemmas\cite{1}

\vspace{.333cm}

\noindent {\it - If $\gamma(q)\equiv{Aq+B\over Cq+D}$, then, up to
a coboundary term, Eq.}(\ref{inhomtrans}) {\it implies}
\begin{equation} (\gamma(q^a);q^b)=(q^a;q^b), \qquad
(q^a;\gamma(q^b))=\left(\partial_{q^b}\gamma(q^b)\right)^{-2}(q^a;q^b).
\label{oiqwdh}\end{equation}

\vspace{.333cm}

\noindent {\it - If $q^a=q^b+\epsilon^{ab}(q^b)$, the unique
solution of Eq.}(\ref{inhomtrans}){\it, depending only on the
first and higher derivatives of $q^a$, is}
\begin{equation}
(q^a;q^b)=c_1{\epsilon^{ab}}'''(q^b)+\mathcal{O}^{ab}(\epsilon^2),\qquad
c_1\ne0. \label{uv}\end{equation}

\noindent Let us review the proof of the second lemma and theorem
1. Since $(q^a;q^b)$ should depend only on $\partial_{q^b}^kq^a$,
$k\geq1$, we have
\begin{equation} (q+\epsilon f(q);q)=c_1\epsilon
f^{(k)}(q)+\mathcal{O}(\epsilon^2),
\label{preliminare1}\end{equation} where $q^a=q+\epsilon f(q)$,
$q\equiv q^b$ and $f^{(k)}\equiv\partial_q^kf$, $k\geq1$. Note
that by (\ref{oiqwdh})
\begin{equation} (Aq+\epsilon Af(q);Aq)=(q+\epsilon
f(q);Aq)=A^{-2}(q+\epsilon f(q);q),
\label{preliminare2}\end{equation} on the other hand, setting
$F(Aq)=Af(q)$, by (\ref{preliminare1}) $(Aq+\epsilon
Af(q);Aq)=(Aq+\epsilon F(Aq);Aq)=
c_1\epsilon\partial_{Aq}^kF(Aq)=A^{1-k}c_1\epsilon f^{(k)}(q)$, so
that $k=3$. One sees that $(Aq+\epsilon Af(q);Aq)$ at order
$\epsilon^n$ is a sum of terms of the form $c_{i_1\ldots
i_n}\partial_{Aq}^{i_1}\epsilon F(Aq)\cdot\cdot\cdot
\partial_{Aq}^{i_n}\epsilon F(Aq)=c_{i_1\ldots i_n}\epsilon^nA^{n-\sum i_k}
f^{(i_1)}(q)\cdot\cdot\cdot f^{(i_n)}(q)$, and by
(\ref{preliminare2}) $\sum_{k=1}^ni_k=n+2$. On the other hand,
since $(q^a;q^b)$ depends only on $\partial_{q^b}^kq^a$, $k\geq1$,
we have $i_k\geq 1$, $k\in[1,n]$, so that either $i_k=3$, $i_j=1$,
$j\in[1,n]$ $j\ne k$, or $i_k=i_j=2$, $i_l=1$, $l\in[1,n]$, $l\ne
k,\,l\ne j$. Hence
\begin{equation} (q+\epsilon
f(q);q)=\sum_{n=1}^\infty\epsilon^n\left(c_nf^{(3)}
f^{(1)^{n-1}}+d_nf^{(2)^2}f^{(1)^{n-2}}\right),\qquad d_1=0.
\label{preliminare4}\end{equation} Inserting the expansion
(\ref{preliminare4}) in (\ref{inhomtrans}), we obtain
\begin{equation} c_n=(-1)^{n-1}c_1,\qquad\qquad
d_n={3\over2}(-1)^{n-1}(n-1)c_1,
\label{cenneedennexf}\end{equation} which are the coefficients in
the power expansion of $c_1\{q+\epsilon f(q),q\}$.

In deriving the equivalence of states we considered the case of
one--particle states with identical masses.\cite{1} The
generalization to the case with different masses is
straightforward. In particular, the right hand side of
Eq.(\ref{ladoppiaww}) gets multiplied by $m_b/m_a$, so that the
cocycle condition becomes
\begin{equation}
m_a(q^a;q^c)=m_a\left(\partial_{q^c}q^b\right)^2(q^a;q^b)+m_b(q^b;q^c),
\label{inhomtransbbb}\end{equation} explicitly showing that the
mass appears in the denominator and that it refers to the label in
the first entry of $(\cdot\,;\cdot)$, that is
\begin{equation}
(q^a;q^b)=-{\hbar^2\over 4m_a}\{q^a;q^b\},
\label{bella}\end{equation} from which the QSHJE (\ref{1Q})
follows almost immediately.\cite{1}

The EP leads to the introduction of length scales,\cite{1,2} a
fact related to the nontriviality of the quantum potential, even
in the case of $\mathcal{W}^0$. We note that also $\mathcal{S}_0$,
as follows by the EP, is never trivial, in particular
\begin{equation}\mathcal{S}_0\ne cnst,\qquad
\forall\mathcal{W}\in{\bf H}.
\label{assolutamentebasilare}\end{equation}

The QSHJE (\ref{1Q}), first investigated by Floyd in a series of
important papers,\cite{Floyd} has been recently studied and
reviewed by several authors.\cite{vari,reviews}

\section{The two--particle model}
The real solution of the QSHJE (\ref{1Q}) is
$e^{{2i\over\hbar}\mathcal{S}_0\{\delta\}}=e^{i\alpha}
{w+i\bar\ell\over w-i\ell}$, $\delta\equiv\{\alpha,\ell\}$, with
$\alpha\in{\bf R}$ and $\ell=\ell_1+i\ell_2$, $\ell_1\neq0$, are
integration constants. The condition $\ell_1\neq0$ is necessary
for $\mathcal{S}_0$ and the quantum potential $Q$ to be
well--defined.

The formulation has a manifest duality between real pairs of
linearly independent solutions,\cite{1} a property strictly
related to Legendre duality.\cite{140} Whereas in the standard
approach one usually considers only one solution of the SE, {\it
i.e.} the wave--function itself, in our formulation the relevant
formulas contain the linear combination $\psi^D+i\ell \psi$. Since
$\ell_1\neq0$, $\psi^D$ and $\psi$ appears always in pair. So,
Legendre duality, nontriviality of $\mathcal{S}_0$ and $Q$ are
deeply related features which are direct consequences of the EP.
In turn, these properties imply the appearance of fundamental
constants such as the Planck length.\cite{1,2} The simplest way to
see this is to consider the SE in the trivial case, that is
$\partial_{q^0}^2\psi=0$, so that $\psi^D=q^0$, $\psi=1$ and the
typical combination reads $q^0+i\ell_0$, implying that
$\ell_0\equiv\ell$ should have the dimension of a length. The fact
that $\ell$ has the dimension of a length is true for any state.
Since $\ell_0$ appears in the QSHJE with $\mathcal{W}^0\equiv0$,
the system does not provide any dimensional quantity, so that we
have to introduce some fundamental length. The appearance of
fundamental constants also arises in considering the limits
$\hbar\to0$ and $E\to0$ in the case of a free particle.\cite{1}
So, for example, a consistent expression for the quantum potential
associated to the trivial state $\mathcal{W}^0$, which vanishes as
$\hbar\to0$, is
\begin{equation} Q^0={\hbar\over
4m}\{\mathcal{S}_0^0,q^0\}=-{\hbar^3 G
\over2mc^3}{1\over|q^0-i\lambda_p|^4},
\label{oix9ui87}\end{equation} where $\lambda_p=\sqrt{\hbar
G/c^3}$ is the Planck length. However, in considering the
classical limit of the reduced action one should include the
gravitational contribution. So, for example, it is clear that also
at the classical level the reduced action for a pair of ``free"
particles should include the Newton potential. This may be related
to the above mentioned appearance of fundamental constants in the
QSHJE.\cite{1,2} Related to this is the Floyd observation that in
the classical limit there is a residual indeterminacy depending on
the integration constants.\cite{Floyd} Thus we see that the
classical limit may in fact lead to some effect which is of
quantum origin even if $\hbar$ does not appear explicitly. We also
note that, in principle, the Planck constant may appear in
macroscopic phenomena. This indicates that it is worth studying
the structure of the quantum potential also at large scales.

It seems that the fundamental properties of $Q$ have not yet fully
been investigated because the usual solutions one finds are
essentially trivial. This is due to a clearly unsatisfactory
identification, that may lead to some inconsistency, of
$Re^{{i\over\hbar}\mathcal{S}_0}$ with the wave--function. As
noticed by Floyd,\cite{Floyd} if $Re^{{i\over\hbar}\mathcal{S}_0}$
solves the SE, then the wave--function will in general have the
form
$R(Ae^{-{i\over\hbar}\mathcal{S}_0}+Be^{{i\over\hbar}\mathcal{S}_0})$.
This simple remark has important consequences. So, for example,
note that a real wave--function, such as the one for bound states,
simply implies $|A|=|B|$ rather than $\mathcal{S}_0=0$. As
Einstein noticed in a letter to Bohm, the latter would imply that
a quantum particle in a box is at rest and starts moving in the
classical limit. Therefore, besides the mathematical consistency,
identifying the wave--function with
$Re^{{i\over\hbar}\mathcal{S}_0}$ cannot in general lead to a
quantum analog of the reduced action. This change in the
definition of $\mathcal{S}_0$ implies a new view of $Q$ which
needs to be further investigated. In this respect we note that $Q$
provides particle's response to an external perturbation. For
example, in the case of tunnelling, where according to the
standard definition $\mathcal{S}_0$ would be vanishing, the
attractive nature of $Q$ guarantees the reality of the conjugate
momentum.\cite{1} As a consequence, the role of this intrinsic
energy, which is a property of all forms of matter, should
manifest itself through effective interactions depending on the
above fundamental constants.

It is therefore natural to consider the so called {\it
two--particle model},\cite{2} consisting of two free particles in
the three dimensional space. This provides a simple physical model
to investigate the structure of the interaction provided by $Q$.
The QSHJE decomposes in equations for the center of mass and for
the relative motion. The latter is the QSHJE
\begin{equation} {1\over 2m}(\nabla\mathcal{S}_0)^2-E-{\hbar^2\over
2m}{\Delta R\over R}=0, \qquad \nabla
\cdot(R^2\nabla\mathcal{S}_0)=0,\label{QSHJEHD}\end{equation}
where $r=r_1-r_2$ and $m={m_1m_2 \over m_1+m_2}$. Due to the
quantum potential the QSHJE has solutions in which the relative
motion is not free as in the classical case. Also note that the
quantum potential is negative definite. Since
$\psi=Re^{{i\over\hbar}\mathcal{S}_0}$ is a solution of the SE, we
have $\mathcal{S}_0={\hbar\over 2i}\ln(\psi/\overline\psi)$, so
that
\begin{equation} (\nabla\mathcal{S}_0)^2=-{\hbar^2\over
4|\psi|^4}(\overline \psi\nabla\psi -\psi\nabla\overline
\psi)^2,\label{iug}\end{equation} where
$\psi=\sum_{l=0}^\infty\sum_{m=-l}^l\sum_{j=1}^2 c_{lmj}
R_{klj}(r)Y_{lm}(\theta,\phi)$, with $Y_{lm}$ the spherical
harmonics and $R_{klj}$ the solutions of the radial part of the
SE.\cite{2}

As $m=l\to\infty$ $P^l_l\propto\sin^l\theta$ vanishes unless
$\theta=\pi/2$, and the motion is on a plane as in the classical
orbits. However, since $\lim_{l\to\infty}\partial_\theta
P^l_l(\cos\theta)=0$, we have $\lim_{m\sim
l\to\infty}\partial_\theta P^m_l(\cos\theta)=0$. Thus, considering
solutions with $c_{lmj}\neq0$ only for sufficiently large $m$ and
$l$, we have
\begin{equation} \nabla\psi=
\sum_{\{lmj\}}\left(c_{lmj}R_{klj}'Y_{lm}, 0, {i\over
r}c_{lmj}R_{klj}mY_{lm}\right).\label{oieufgh}\end{equation}

Depending on the coefficients $c_{lmj}$,
$(\nabla\mathcal{S}_0)^2/2m$ may contain nontrivial terms which do
not cancel as $\hbar\to0$. These may arise as a deformation of the
classical kinetic term, which includes the centrifugal term. The
$c_{lmj}$, which may depend on fundamental constants, fix the
structure of the possible interaction in the two--particle model.
The $c_{lmj}$ may be related to some boundary conditions implied
by the geometry and the matter content of the three--dimensional
space. This would relate the fundamental constants to possible
collective effects\cite{2} which may depend on cosmological
aspects.

\section*{Acknowledgments} It is a pleasure to thank G. Bertoldi
and A.E. Faraggi for collaboration and E.R. Floyd, J.M Isidro, L.
Mazzucato, P. Pasti and M. Tonin for stimulating discussions. Work
partially supported by the European Community's Human Potential
Programme under contract HPRN-CT-2000-00131 Quantum Spacetime.

\end{document}